\documentclass[aps,twocolumn,superscriptaddress]{revtex4-1}
\usepackage{graphicx}
\usepackage{bm}
\usepackage{amssymb}
\usepackage{amsmath}
\usepackage{epstopdf}

\begin{document}

\title{Understanding the magnetism in noncentrosymmetric CeIrGe$_3$: Muon spin relaxation and neutron scattering studies }

\author{V.\ K.\ Anand}
\altaffiliation{vivekkranand@gmail.com}
\affiliation{ISIS Facility, Rutherford Appleton Laboratory, Chilton, Didcot, Oxon, OX11 0QX, United Kingdom}
\affiliation{\mbox{Helmholtz-Zentrum Berlin f\"{u}r Materialien und Energie GmbH, Hahn-Meitner Platz 1, D-14109 Berlin, Germany}}
\author{A.\ D.\ Hillier}
\affiliation{ISIS Facility, Rutherford Appleton Laboratory, Chilton, Didcot, Oxon, OX11 0QX, United Kingdom}
\author{D.\ T.\ Adroja}
\altaffiliation{devashibhai.adroja@stfc.ac.uk}
\affiliation{ISIS Facility, Rutherford Appleton Laboratory, Chilton, Didcot, Oxon, OX11 0QX, United Kingdom}
\affiliation{\mbox{Highly Correlated Matter Research Group, Physics Department, University of Johannesburg, P.O. Box 524,} Auckland Park 2006, South Africa}
\author{D.~D.~Khalyavin}
\author{P.\ Manuel}
\affiliation{ISIS Facility, Rutherford Appleton Laboratory, Chilton, Didcot, Oxon, OX11 0QX, United Kingdom}
\author{G.\ Andre}
\affiliation{Laboratoire Leon Brillouin (CEA-CNRS), CEA/Saclay, F-91191 Gif-sur-Yvette, France}
\author{S.\ Rols}
\author{M.\ M.\ Koza}
\affiliation{Institut Laue-Langevin, 71 avenue des Martyrs, F-38042 Grenoble Cedex 9, France}

\date{\today}

\begin{abstract}
The magnetic properties of a pressure induced noncentrosymmetric heavy-fermion superconductor CeIrGe$_3$ have been investigated by muon spin relaxation ($\mu$SR), powder neutron diffraction (ND) and inelastic neutron scattering (INS) techniques at ambient pressure. For completeness we have also measured the ac magnetic susceptibility $\chi_{\rm ac}(T)$, dc magnetic susceptibility $\chi(T)$, dc isothermal magnetization $M(H)$ and heat capacity $C_{\rm p}(T,H)$ down to 2~K\@. CeIrGe$_{3}$ is known to exhibit pressure induced superconductivity ($T_{\rm c}\approx 1.5$~K) at a pressure of 20~GPa and antiferromagnetic ordering at 8.7~K, 4.7~K and 0.7~K at ambient pressure. Our $\chi_{\rm ac}(T)$ and $\chi(T)$ data show an additional anomaly near 6.2~K which is also captured in $C_{\rm p}(T)$ data. From $\chi_{\rm ac}(T)$, $\chi(T)$ and $C_{\rm p}(T)$ measurements we infer three antiferromagnetic transitions above 2~K at $T_{\rm N1}= 8.5$~K, $T_{\rm N2}= 6.0$~K  and $T_{\rm N3}= 4.6$~K\@. Our $\mu$SR study also confirms the presence of three transitions through the observation of one frequency for $T_{\rm N2} < T\leq T_{\rm N1}$, two frequencies for $T_{\rm N3} < T\leq T_{\rm N2}$ and three frequencies for $T\leq T_{\rm N3}$ in the oscillatory asymmetry. The ND data reveal an incommensurate nature of the magnetic ordering at $T=7$~K with the propagation vector {\bf k} = (0,0,0.688(3)),  and a commensurate magnetic structure at $T=1.5$~K with the propagation vector locked to the value {\bf k} = (0,\,0,\,2/3) and magnetic moments oriented along the $c$ axis. The commensurate structure couples a macroscopic ferromagnetic component, resulting in a strong dependence of the lock-in transition temperature on external magnetic field. The INS data show two well defined crystal electric field (CEF) excitations arising from the CEF-split Kramers doublet ground state of Ce$^{3+}$. The CEF energy levels scheme and wavefunctions have been determined. The ND and INS results together suggest that the anisotropic magnetic exchange are playing an important role in the magnetism of CeIrGe$_3$.
\end{abstract}

\maketitle

\section{INTRODUCTION}

Recently, a new class of materials that lack an inversion symmetry in their crystal structure have attracted lots of attention for their exotic superconductivity on account of an antisymmetric spin-orbit coupling (ASOC) \cite{Bauer2012}. In these \emph{noncentrosymmetric} superconductors (NCSs) the ASOC removes the spin degeneracy of conduction band electrons, implying that the spin and orbital parts of the Cooper pair wave function cannot be treated independently. Therefore parity is no longer a good quantum number and a parity mixing of spin-singlet and spin-triplet states occurs \cite{Edelstein, Gorkov, Samokhin, Frigeri,Fujimoto2007, Saxena2004}. This behavior of noncentrosymmetric superconductors is very different from that of \emph{centrosymmetric} superconductors which have degenerate conduction band irrespective of the strength of spin-orbit coupling, the Cooper pair wave function of the latter thus consists of a pure spin-singlet ({\it s}-wave) or spin-triplet ({\it p}-wave) pairing. 

The noncentrosymmetric heavy fermion superconductor CePt$_{3}$Si crystallizing in a primitive tetragonal structure (space group $P4\,mm$) has been found to exibit many unusual superconducting properties because of the presence of Rashba-type ASOC as a result of the lack of an inversion symmetry in its crystal structure \cite{Bauer2004, Bauer2005, Bauer2007}. Interestingly, superconductivity ($T_{\rm c}=0.75$~K) in this compound coexists with an antiferromagnetic (AFM) order ($T_{\rm N} =2.2$~K) \cite{Bauer2004}. The upper critical field $H_{\rm c2} \approx 5$~T of CePt$_{3}$Si is much higher than the expected Pauli paramagnetic limiting field $H_{\rm P} \sim 1$~T for spin-singlet pairing. This provides evidence for the order parameter to have a mixed spin singlet-triplet state  \cite{Bauer2004,Fujimoto2007}. 

After the report of noncentrosymmetric superconductivity in CePt$_{3}$Si, a number of noncentrosymmetric materials have been reported to exhibit interesting superconducting properties \cite{Bauer2012}. Of particular interests are the compounds Ce$TX_3$ ($T =$ transition metal and $X=$ Si, Ge) having a BaNiSn$_{3}$-type noncentrosymmetric tetragonal structure (space group $I4\,mm$) which lack a mirror plane symmetry along the $c$-axis and host a Rashba-type ASOC. Among noncentrosymmetric Ce$TX_3$ the noncentrosymmetric heavy fermion superconductors CeRhSi$_{3}$, CeIrSi$_{3}$, CeCoGe$_{3}$ and  CeIrGe$_{3}$, all of which exhibit a long-range antiferromagnetic ordering at ambient pressure, have been found to show superconductivity under applied pressure with $T_{\rm c}= 0.7$--1.6~K \cite{Muro1998,Kimura2005,Kimura2007,Sugitani2006, Knebel2009, Settai2007, Honda2010, Okuda2007, Thamizhavel2005, Kawai2008}. However, their nonmagnetic analogs such as LaRhSi$_3$, LaIrSi$_{3}$, LaPdSi$_{3}$ and LaPtSi$_{3}$ exhibit superconductivity at ambient pressure with $T_{\rm c}=0.7$--2.7~K  \cite{Okuda2007, Anand2011a, Anand2014a, Anand2015a, Smidman2014}. Valence fluctuating CeCoSi$_{3}$ on the other hand is reported to exhibit superconductivity at ambient pressure (sample dependent $T_{\rm c}= 0.7$--1.4~K) \cite{Haen1985, Iwamoto1995}. While Ce-based NCSs exhibit exotic superconducting ground state, most of the nonmagnetic La-based NCSs behave like conventional $s$-wave superconductors. Thus the magnetic fluctuations seem to be important for the exotic behavior of NCSs. The relationship between superconductivity and the lack of inversion symmetry as well as the role of magnetic fluctuations in noncentrosymmetric $f$-electron systems is still puzzling.

At ambient pressure, CeRhSi$_{3}$ exhibits antiferromagnetic ordering below $T_{\rm N}= 1.6$~K,  superconductivity appears at pressures $p>1.2$~GPa ($T_{\rm c} \sim1$~K) that coexists with AFM. A large $H_{\rm c2} \approx 7$~T ($H_{\rm P} \sim 1.6$~T) reflects the influence of ASOC on the superconducting state \cite{Kimura2005,Kimura2007}. CeIrSi$_{3}$ exhibits antiferromagnetic ordering below $T_{\rm N}= 5$~K at ambient pressure and superconductivity appears at $p\approx 2.5$~GPa ($T_{\rm c}\approx 1.6$~K) after the AFM order is suppressed \cite{Sugitani2006, Okuda2007}. The anisotropic $H_{\rm c2} \approx 30$~T for $H\parallel c$ and $H_{\rm c2} \approx 9.5$~T for $H\perp c$ is much larger than $H_{\rm P} \approx 3$~T, indicating a mixed pairing state \cite{Sugitani2006, Okuda2007}. CeCoGe$_{3}$ exhibits three successive AFM transitions at $T_{\rm N1}= 21$~K, $T_{\rm N2}= 12$~K and  $T_{\rm N3}= 8$~K at ambient pressure and become superconducting at  $p > 5.4$~GPa with $T_{\rm c}\approx 0.7$~K \cite{Kawai2008,Thamizhavel2005,Knebel2009, Settai2007}.

CeIrGe$_{3}$ which is the subject of the present work is reported to order antiferromagnetically, showing three transitions at $T_{\rm N1}= 8.7$~K, $T_{\rm N2}= 4.7$~K and  $T_{\rm N3}= 0.7$~K \cite{Muro1998}. The pressure study on CeIrGe$_{3}$ have revealed that $T_{\rm N1}=8.7$~K remains nearly constant, but $T_{\rm N2}=4.7$~K increases with pressure until it merges with $T_{\rm N1}$ at 4~GPa, superconductivity appears at pressures above 20 GPa ($T_{\rm c}\approx 1.5$~K) coexisting with AFM up to 22~GPa above which AFM order is suppressed completely. At 24 GPa superconductivity is accompanied with non-Fermi liquid behavior \cite{Kawai2008, Honda2010}. A large $H_{\rm c2} \approx 10$~T ($H_{\rm P} \sim 3$~T) for $H\parallel c$ was found as is commonly seen in other NCSs.  CeRhGe$_{3}$ on the other hand, also with antiferromagnetic ground state, does not show pressure induced superconductivity, the $T_{\rm N1}$ increases with increasing pressure and reaches 21.3~K at 8.0~GPa from 14.6~K at ambient pressure \cite{Muro1998,Kawai2008}. 

In our efforts to understand the magnetism and the role of single-ion anisotropy arising from crystal field in $RTX_3$ system \cite{Anand2011b, Anand2011c, Anand2012a, Adroja2012a, Adroja2012b, Anand2013, Anand2014b, Anand2015b, Anand2016, Adroja2015} recently some of us have  investigated the magnetic properties of BaNiSn$_{3}$-type noncentrosymmetric materials CeRhGe$_{3}$ \cite{Hillier2012}, CeCoGe$_{3}$ \cite{Smidman2013} and CeRuSi$_{3}$ \cite{Smidman2015} using neutron scattering and muon spin relaxation ($\mu$SR) techniques. Our $\mu$SR study on heavy fermion CeRhGe$_{3}$ revealed clear frequency oscillations indicating two  AFM transitions at $T_{\rm N1}= 14.5$~K and  $T_{\rm N2}= 7$~K  \cite{Hillier2012}. The magnetic structure determination by powder neutron diffraction (ND) revealed a spin-density-wave-type magnetic ordering of Ce$^{3+}$ moments [ordered moment of 0.45(9)\,$\mu_{\rm B}$] represented by propagation vector ${\bf k} = (0\, 0\, \frac{3}{4})$  along the $c$-axis \cite{Hillier2012}. The observed $c$-axis moment direction differs from the expected single-ion $ab$-plane CEF anisotropy, which is ascribed to the presence of two-ion anisotropic exhange interaction.   The inelastic neutron scattering (INS) data revealed the presence of two well-defined crystal field (CEF) excitations at 7.5~meV and 18~meV. The INS data indicated a local moment magnetism in CeRhGe$_{3}$ which is thought to be responsible for the absence of pressure induced superconductivity in this compound \cite{Hillier2012}. 

Our $\mu$SR investigations on CeCoGe$_{3}$ revealed clear frequency oscillations associated with AFM orderings \cite{Smidman2013}. As stated above, CeCoGe$_{3}$ exhibits three transitions at $T_{\rm N1}= 21$~K, $T_{\rm N2}= 12$~K and  $T_{\rm N3}= 8$~K which are further confirmed by ND data.  The single crystal ND data revealed that the three AFM phases of CeCoGe$_{3}$ are characterized by the propagation vectors ${\bf k} = (0\, 0\, \frac{2}{3})$ between $T_{\rm N1}$ and $T_{\rm N2}$, ${\bf k} = (0\, 0\, \frac{5}{8})$ between $T_{\rm N2}$ and $T_{\rm N3}$, and ${\bf k} = (0\, 0\, \frac{1}{2})$ below $T_{\rm N3}$ \cite{Smidman2013}. The magnetic structure turns out to be an equal moment two-up two-down below $T_{\rm N3}$ [ordered moment of 0.405(5)\,$\mu_{\rm B}$] and equal moment two-up one-down above $T_{\rm N2}$ [ordered moment of 0.360(6)\,$\mu_{\rm B}$] \cite{Smidman2013}. The INS data show two well-defined  CEF excitations at 19~meV and 27~meV and present evidence for $c$-$f$ hybridization in CeCoGe$_{3}$ \cite{Smidman2013}.  The INS investigations on valence fluctuating CeRuSi$_{3}$ have shown existence of a hybridization gap \cite{Smidman2015}.

Here we present the results of neutron scattering and $\mu$SR studies on the noncentrosymmetric heavy fermion superconductor CeIrGe$_{3}$. Consistent with the heat capacity and magnetic susceptibility, the $\mu$SR data reveal three magnetic transitions above 2~K at $T_{\rm N1}= 8.5$~K, $T_{\rm N2}= 6.0$~K  and $T_{\rm N3}= 4.6$~K\@.  It should be noted that previous investigations on CeIrGe$_{3}$ by Muro {\it et al}.\ \cite{Muro1998} and Kawai {\it et al}.\ \cite{Kawai2008} could not capture the 6~K transition. Our powder ND data reveal an incommensurate magnetic structure with propagation vector {\bf k} = (0,0,0.688) below $T_{\rm N1}$ which then locks to the commensurate propagation vector {\bf k} = (0,\,0,\,2/3) in the ground state. The commensurate magnetic phase, taking place at $T_{\rm N3}$, couples a macroscopic ferromagnetic component and can be induced by external magnetic filed above this temperature. This results in experimentally observed metamagnetic behaviour and implies magnetic-filed induced lock-in transition. The INS data reflect two well defined excitations of magnetic origin accounted by the crystal field model. 

\section{Experimental details}

The polycrystalline samples of CeIrGe$_{3}$ and LaIrGe$_{3}$ were prepared by arc melting stoichiometric amounts of high purity ($\geq 99.9$\%) Ce, La, Ir and Ge on a water cooled copper hearth in inert argon atmosphere. The resulting ingots were flipped and re-melted several times to ensure homogeneity. The arc melted ingots were annealed for a week at 950~$^\circ$C to improve the phase purity. The sample quality was checked by powder x-ray diffraction which showed the sample to be single phase without any evidence of an impurity phase. The dc magnetization measurements were carried out using a Quantum Design magnetic property measurement system SQUID magnetometer, and the ac magnetization and heat capacity measurements were carried out using a Quantum Design physical properties measurement system. 

The muon spin relaxation measurements were performed at the ISIS facility of Rutherford Appleton Laboratory, U.K.\ using the MuSR spectrometer in longitudinal geometry configuration. The powdered CeIrGe$_{3}$ sample was mounted on a high purity Ag plate using GE varnish which was then covered with a thin layer of mylar film. The sample mount was cooled down to 1.2~K using He-exchange gas in an Oxford Instruments Variox cryostat. The powder neutron diffraction measurements were also performed at the ISIS facility using the time-of-flight diffractometer WISH. The powdered CeIrGe$_{3}$ sample was mounted in a 6~mm vanadium can and cooled down to 1.5~K using a standard He-cryostat. The ND data were collected at three temperatures 1.5~K, 7~K and 10~K, with a long counting time of six hours at each temperature. 

The inelastic neutron scattering measurements were carried out at the Institut Laue-Langevin (ILL), Grenoble, France on the time-of-flight (TOF) spectrometer IN4. The powder CeIrGe$_{3}$ samples were wrapped in a thin Al-foil which was mounted inside a thin-walled cylindrical Al-can that was cooled down to 2 K using a top-loading closed-cycle-refrigerator with He-exchange gas environment. The INS data were collected at 2~K, 10~K and 100~K with neutrons of incident energy $E_i = 67.6$~meV. In order to determine the magnetic scattering of CeIrGe$_{3}$ we also measured the INS response of isostructural nonmagnetic reference compound LaIrGe$_{3}$ that served as a phonon background. The low-energy INS measurements on CeIrGe$_{3}$  were also carried out at ILL using the TOF spectrometer IN6. The low energy INS data were collected at 1.5~K and 10~K with neutrons of $E_i =3.1$~meV.

\begin{figure}
\includegraphics[width=\columnwidth, keepaspectratio]{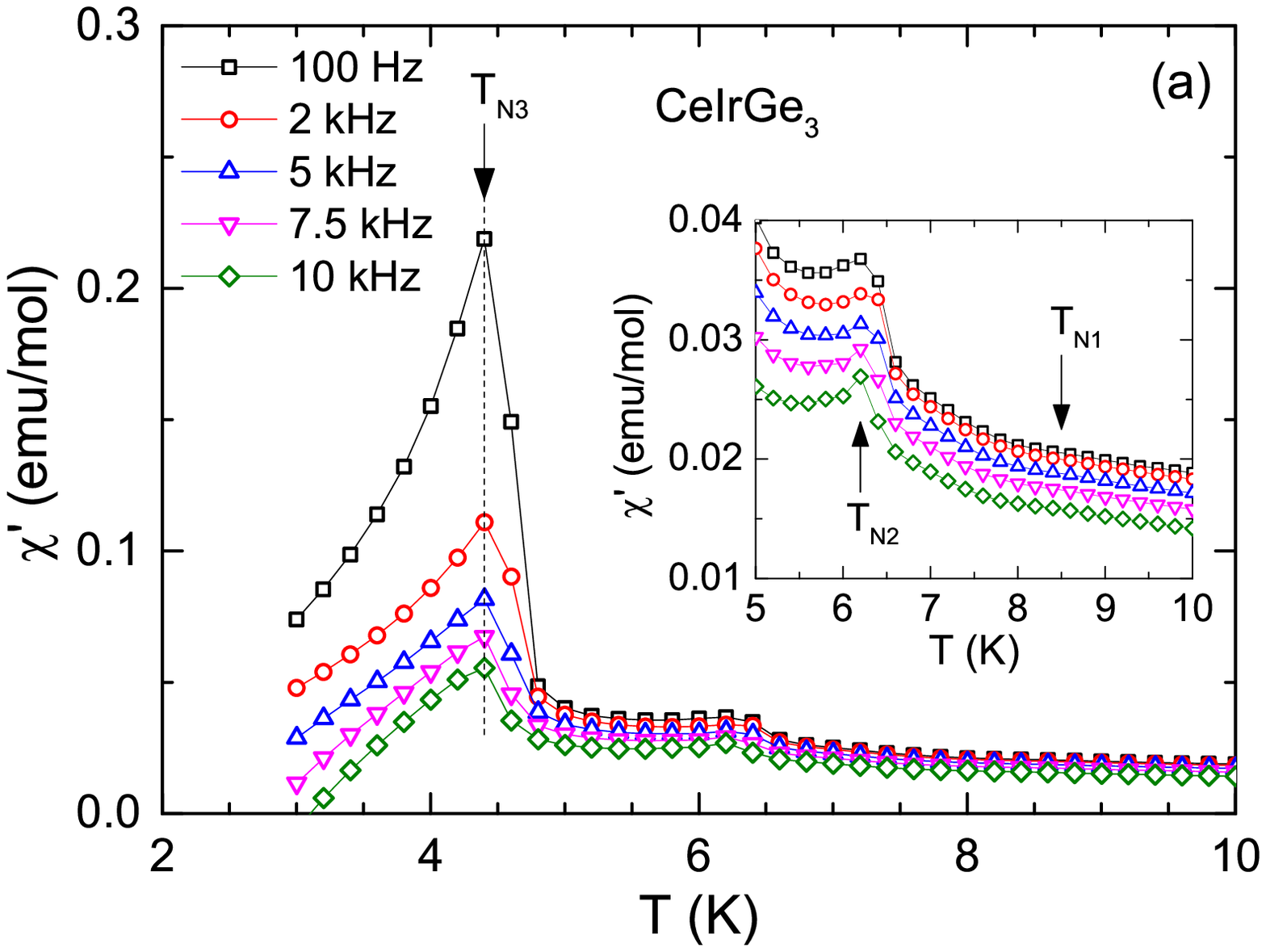}\vspace{0.05in} 
\includegraphics[width=\columnwidth, keepaspectratio]{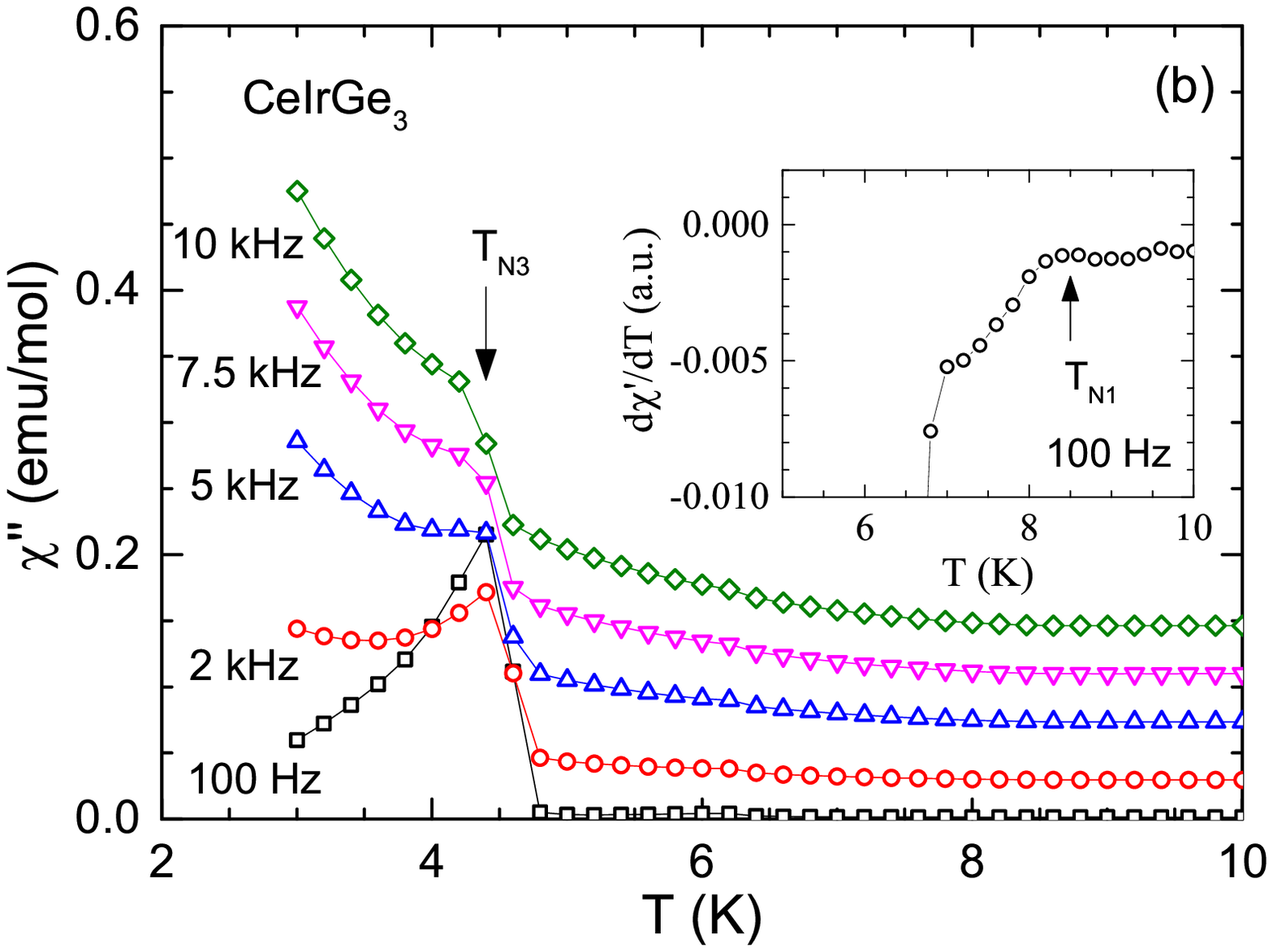}
\caption {(Color online) Temperature $T$ dependence of (a) real $\chi'$ and (b) imaginary $\chi''$ parts of the ac magnetic susceptibility $\chi_{\rm ac}$ of CeIrGe$_{3}$ measured at different frequencies from 100~Hz to 10~kHz in an applied ac magnetic field of 1.0~mT\@. The vertical dotted line in (a) is a guide to the eyes showing the peak positions. Inset in (a) shows an expanded plot of $\chi'(T)$ data. Inset in (b) shows derivative plot ($d\chi'/dT$ vs $T$) of 100~Hz $\chi'(T)$ data showing the weak anomaly near 8.5~K\@. The arrows mark the anomalies associated with $T_{\rm N1}$, $T_{\rm N2}$  and $T_{\rm N3}$.}
\label{fig:Chiac}
\end{figure}

\begin{figure}
\includegraphics[width=\columnwidth, keepaspectratio]{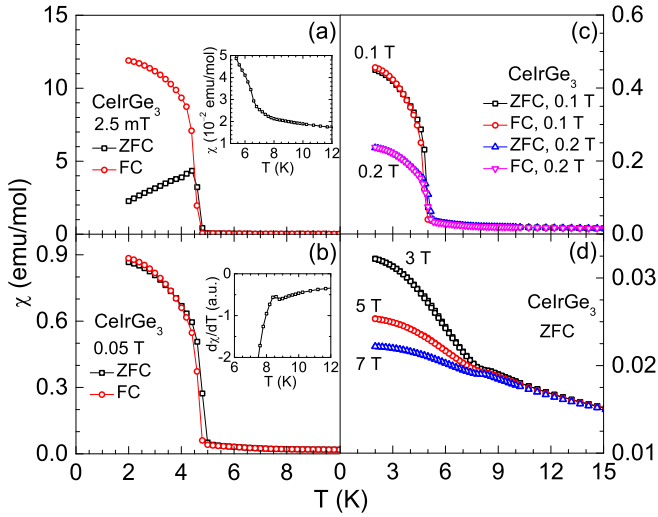}
\caption {(Color online) Temperature $T$ dependence of zero field cooled (ZFC) and field cooled (FC) dc magnetic susceptibility $\chi$ of CeIrGe$_{3}$ in the temperature range 2--15~K measured in magnetic field (a) $H= 2.5$~mT, (b) $H=  0.05$~T, (c) $H= 0.1$ and 0.2~T, and (d) $H= 3$--7~T\@. Inset in (a) shows an expanded scale view of 2.5~mT ZFC $\chi(T)$ data, and inset in (b) shows the derivative plot ($d\chi/dT$  vs $T$) of 0.05~T ZFC $\chi(T)$ data (the  $d\chi/dT$ can be converted to emu/mol\,K by mutiplying with a factor of 0.002).}
\label{fig:Chi-LowH}
\end{figure}

\section{\label{Sec:Magnetization} \lowercase{ac} and \lowercase{dc} Magnetic Susceptibility and Magnetization}

Figure~\ref{fig:Chiac} shows the temperature $T$ dependence of the real $\chi'$ and imaginary $\chi''$ parts of the ac susceptibility $\chi_{\rm ac}$ of CeIrGe$_{3}$ for different frequencies ($100~{\rm Hz} \leq \nu \leq 10$~kHz) measured with an ac magnetic field of 1.0~mT. Both $\chi'(T)$ and $\chi''(T)$ show a well pronounced anomaly near 4.5~K. As marked by the vertical dotted line in Fig.~\ref{fig:Chiac}(a)  the temperature position of the $\chi'(T)$ anomaly (peak) is frequency independent over entire range of measured frequency $100~{\rm Hz} \leq \nu \leq 10$~kHz, though the magnitude of $\chi'(T)$ and $\chi''(T)$ are $\nu$ dependent. The $\chi''(T)$ also show an almost frequency independent anomaly, though the $T$ dependence of $\chi''$ for $T< 4.5$~K is significantly modified by increasing $\nu$ [Fig.~\ref{fig:Chiac}(b)]. While at low frequency (100~Hz) $\chi''$ decreases with decreasing $T$ at $T< 4.5$~K, at $\nu\geq 5$~kHz the $\chi''$ increases as $T$ is lowered. The magnitude of $\chi''$ increases with increasing $\nu$ which is opposite to the observation for $\chi'$ which decreases with increasing $\nu$.    

We also see another weak anomaly in $\chi'(T)$ near 6.2~K which is better seen from the expanded scale plot in the inset of Fig.~\ref{fig:Chiac}(a), though this anomaly is not visible in $\chi''(T)$. The position of the 6.2~K anomaly is also $\nu$ independent. Previous investigations by Muro {\it et al}.\ \cite{Muro1998} and Kawai {\it et al}.\ \cite{Kawai2008} did not report any anomaly near 6~K. As discussed below the dc susceptibility and heat capacity data also show anomaly near 6~K which reflects the intrinsic nature of this anomaly. The frequency independent behavior of the 4.5~K and 6.2~K anomalies suggests that they are not related to spin-freezing, instead they represent the long range magnetic phase transitions. Further, a weak change in slope, better visualized in derivative plot ($d\chi'/dT$ vs $T$) shown in the inset of Fig.~\ref{fig:Chiac}(b), mark another anomaly near 8.5~K where a well pronounced anomaly is seen in the heat capacity data.

\begin{figure}
\includegraphics[width=\columnwidth, keepaspectratio]{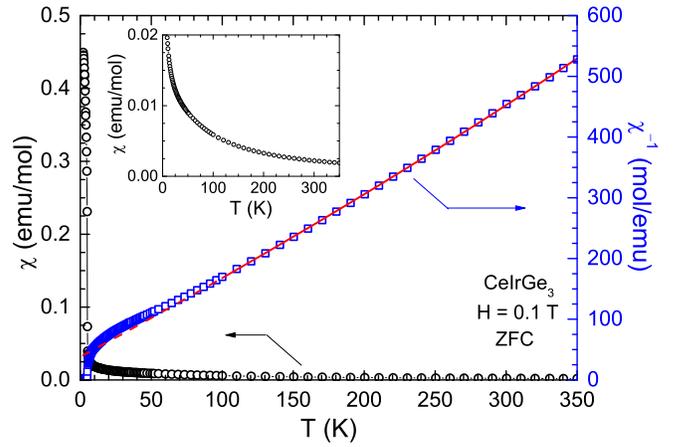}
\caption {(Color online) Temperature $T$ dependence of zero field cooled dc magnetic susceptibility $\chi$ (left ordinate) of CeIrGe$_{3}$ and its inverse $\chi^{-1}$ (right ordinate) in the temperature range 2--350 K measured in a magnetic field $H =0.1$~T\@. The solid line represents the fit to Curie-Weiss law over $100~{\rm K} \leq T \leq 350$~K and the dashed line is an extrapolation. Inset: An expanded scale view of $\chi(T)$ data.}
\label{fig:Chi}
\end{figure}

\begin{figure}
\includegraphics[width=\columnwidth, keepaspectratio]{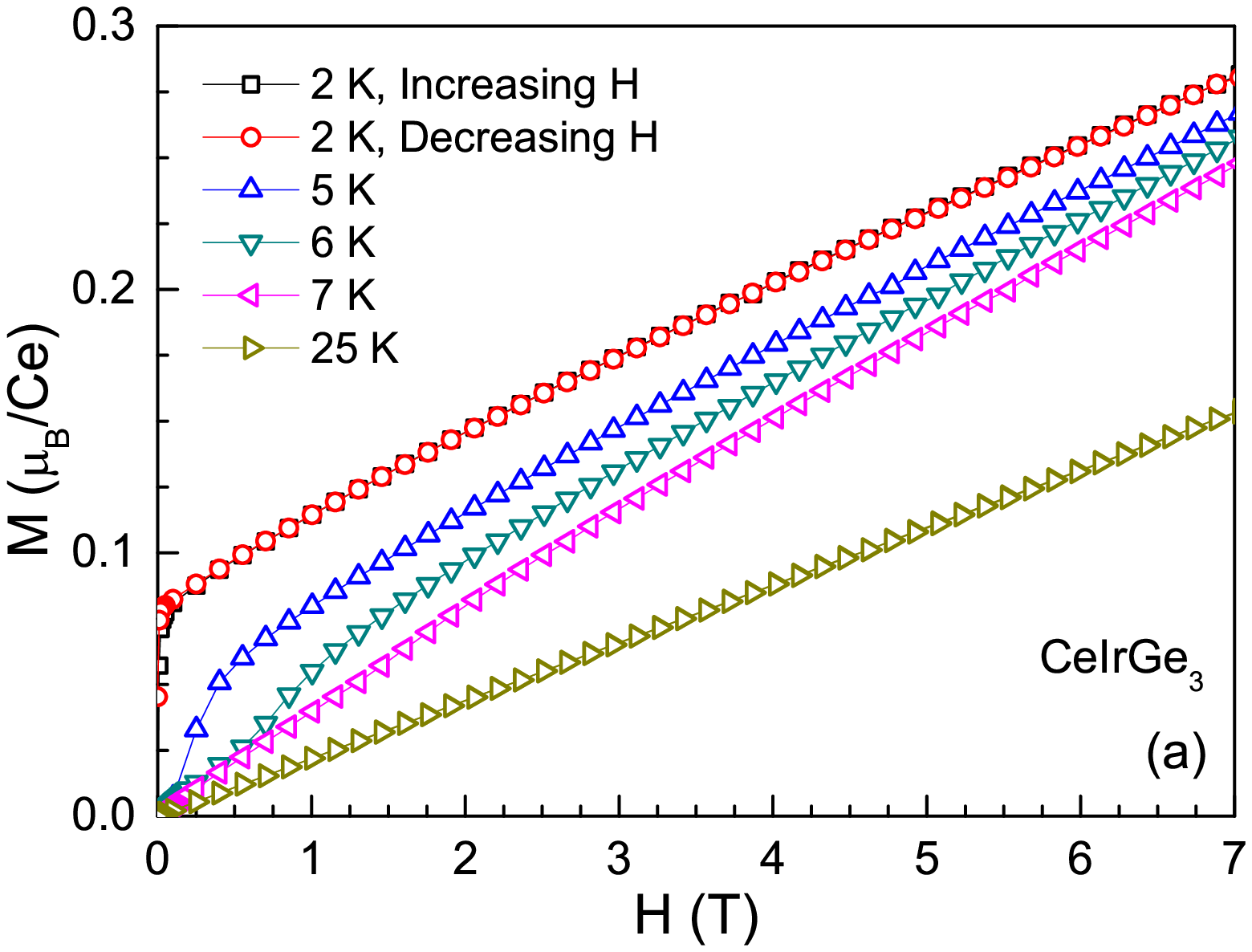}
\includegraphics[width=\columnwidth, keepaspectratio]{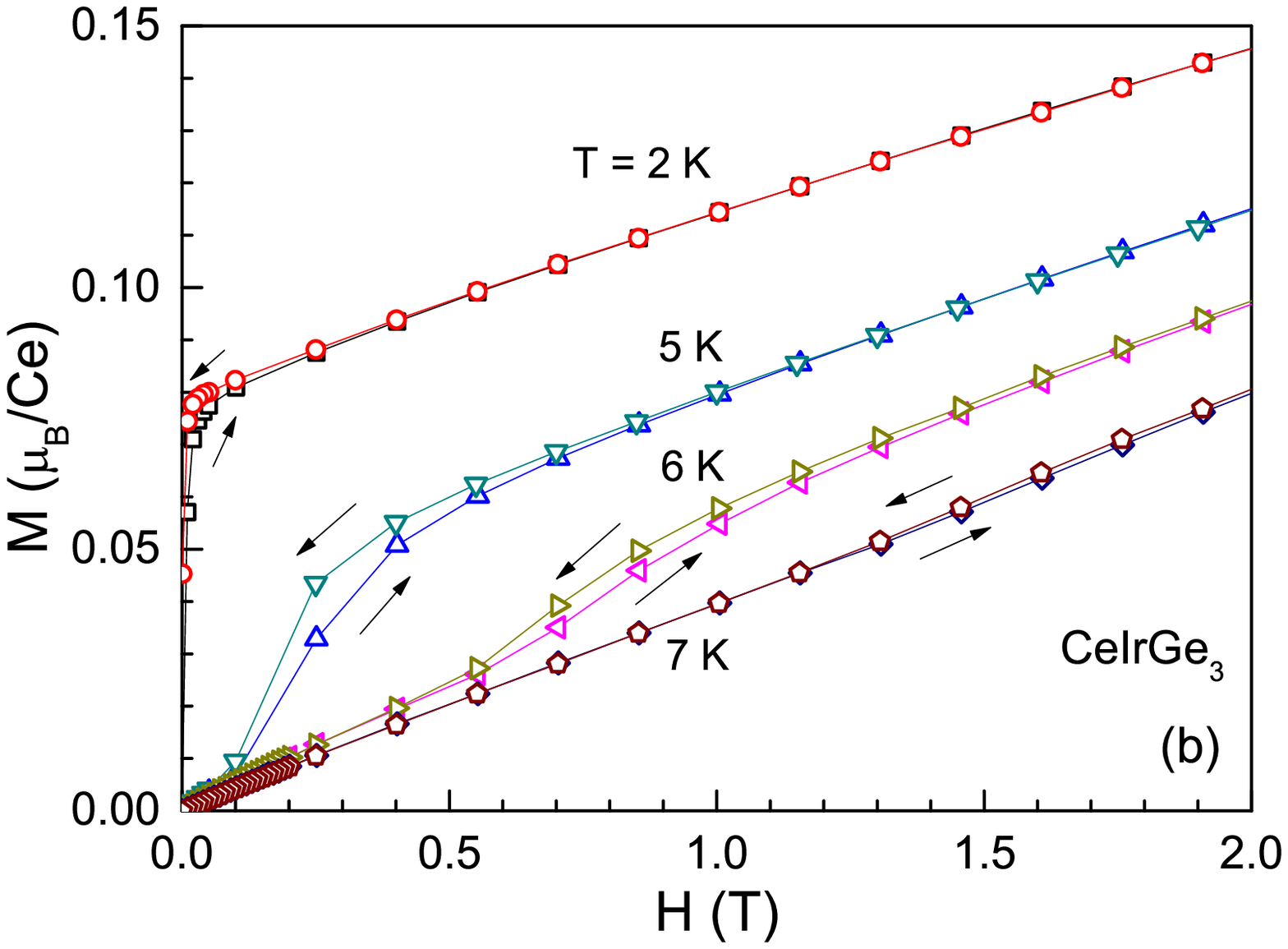}
\caption {(Color online) (a) Magnetic field $H$ dependence of dc isothermal magnetization $M$ of CeIrGe$_{3}$ measured at the indicated temperatures for $H\leq 7$~T\@. (b) $M(H)$ data for $H \leq 2$~T\@. The arrows mark the direction of increasing and decreasing $H$.}
\label{fig:MH}
\end{figure}

Figure~\ref{fig:Chi-LowH} shows the low-$T$ zero-field-cooled (ZFC) and field-cooled (FC) dc magnetic susceptibility $\chi(T)$ of CeIrGe$_{3}$ for different applied magnetic field $H$. As seen from Fig.~\ref{fig:Chi-LowH}(a)  the $\chi(T)$ measured at $H= 2.5$~mT shows a very rapid increase below 4.8~K and peaks near 4.5~K. We also notice a clear splitting between the ZFC and FC data. While the 4.5~K anomaly is present even at higher fields, the splitting between the ZFC and FC data almost disappears at $H\geq 0.05$~T [Figs.~\ref{fig:Chi-LowH}(b) and \ref{fig:Chi-LowH}(c)]. A weak increase in temperature marking the rapid increase of $\chi$, from 4.8~K at 2.5~mT [Fig.~\ref{fig:Chi-LowH}(a)] to 5.2~K at 0.2~T [Fig.~\ref{fig:Chi-LowH}(c)], is also observed. The relatively large magnitude of $\chi$ below 4.5~K and irreversibility between ZFC and FC $\chi$ suggest for the presence of significant ferromagnetic fluctuations and/or formation of antiferromagnetic domains.

The expanded scale plot of 2.5~mT $\chi(T)$ data [see inset of Fig.~\ref{fig:Chi-LowH}(a)] shows a weak anomaly near 6.2~K, consistent with the anomaly in $\chi_{\rm ac}(T)$ data discussed above. However, the anomaly near 8.5~K is too waek to be visible. As shown in the inset of Fig.~\ref{fig:Chi-LowH}(b), the derivative plot ($d\chi/dT$  vs $T$) shows a clear change in the slope near 8.5~K. At $H= 3$~T [Fig.~\ref{fig:Chi-LowH}(d)] one can clearly see a weak cusp near 8.5~K which shifts towads the lower temperature side with increasing field (e.g., to 8.2~K at 7~T). The 8.5~K anomaly thus is related to the occurence of an antiferromagnetic transtion. The 6.2~K anomaly is very likely related to a spin-reorientation transtion in antiferromagnetic state. 

The ZFC $\chi(T)$ and its inverse $\chi^{-1}(T)$ for the temperature range 2~K~$\leq T\leq$~350~K (measured in $H = 0.1$~T) are shown in Fig.~\ref{fig:Chi}. The paramagnetic state $\chi(T)$ data follow the modified Curie-Weiss behavior, $\chi(T) = \chi_0 + {\mbox C/(T-\theta_{\rm p})}$. The fit of $\chi^{-1}(T)$ data  in the temperature range 100~K~$\leq T\leq$~350~K is shown by the solid red line in Fig.~\ref{fig:Chi}, giving a $T$ independent susceptibility of $\chi_0 = -2.3(2) \times 10^{-4}$ emu/mol, Curie constant $C = 0.81(1)$~emu\,K/mol and Weiss temperature $\theta_{\rm p} = -31(2)$~K\@. The effective moment $\mu_{\rm eff}$ estimated from $C$ is $2.55(1)\, \mu_{\rm B}$/Ce in very good agreement with the theoretically expected value of $2.54 \, \mu_{\rm B}$ for Ce$^{3+}$ ions. The negative value of  $\theta_{\rm p}$ reflects a dominant antiferromagnetic interaction in CeIrGe$_{3}$.

Figure~\ref{fig:MH} shows the isothermal magnetization $M(H)$ of CeIrGe$_{3}$ measured at selected temperatures of 2, 5, 6, 7, and 25~K\@. At 2~K, the $M(H)$ isotherm exhibits a ferromagnetic-like spontaneous magnetization at low-$H$ and then $M$ increases almost linearly with $H$. We also see a very small magnetic hysteresis below 0.05~T in $M(H)$ at 2~K [see Fig.~\ref{fig:MH}(b)]. At 5~K, the $M(H)$ isotherm shows a metamagnetic like feature near 0.2~T and a magnetic hysteresis below 0.8~T\@ [Fig.~\ref{fig:MH}(b)].  The $M(H)$ isotherm at 6~K also shows the metamagnetic feature at relatively higher field but with a narrower hysteresis loop [Fig.~\ref{fig:MH}(b)]. The metamagnetic feature becomes very weak at 7~K and hysteresis loop becomes extremely narrow. In view of the metamagnetic behavior at 5~K and 6~K, the ferromagnetic-like spontaneous behavior of $M(H)$ at 2~K could be understood to be the result of an extremely small critical field of metamagnetic transtion at 2~K. The observation of hysteresis in antiferromagnetic state likely reflects the formation of antiferromagnetic domains \cite{Tanner1979}.

The $M(H)$ isotherms do not saturate up to 7~T, the value of $M \approx  0.28\, \mu_{\rm B}$/Ce at 2~K and 7~T is very low compared to the theoretical value of saturation magnetization $M_{\rm s}$ = 2.14\,$\mu_{\rm B}$/Ce for Ce$^{3+}$ ions ($J=5/2$). The reduction in $M$ can be attributed to the combined effect of Kondo effect and CEF. Our $M(H)$ data are consistent with those reported by Muro {\it et al}.\ \cite{Muro1998} and Kawai {\it et al}.\ \cite{Kawai2008}, however they did not explore the hysteresis in $M(H)$ curves.

\section{\label{Sec:HC} Heat Capacity}

\begin{figure}
\includegraphics[width=\columnwidth]{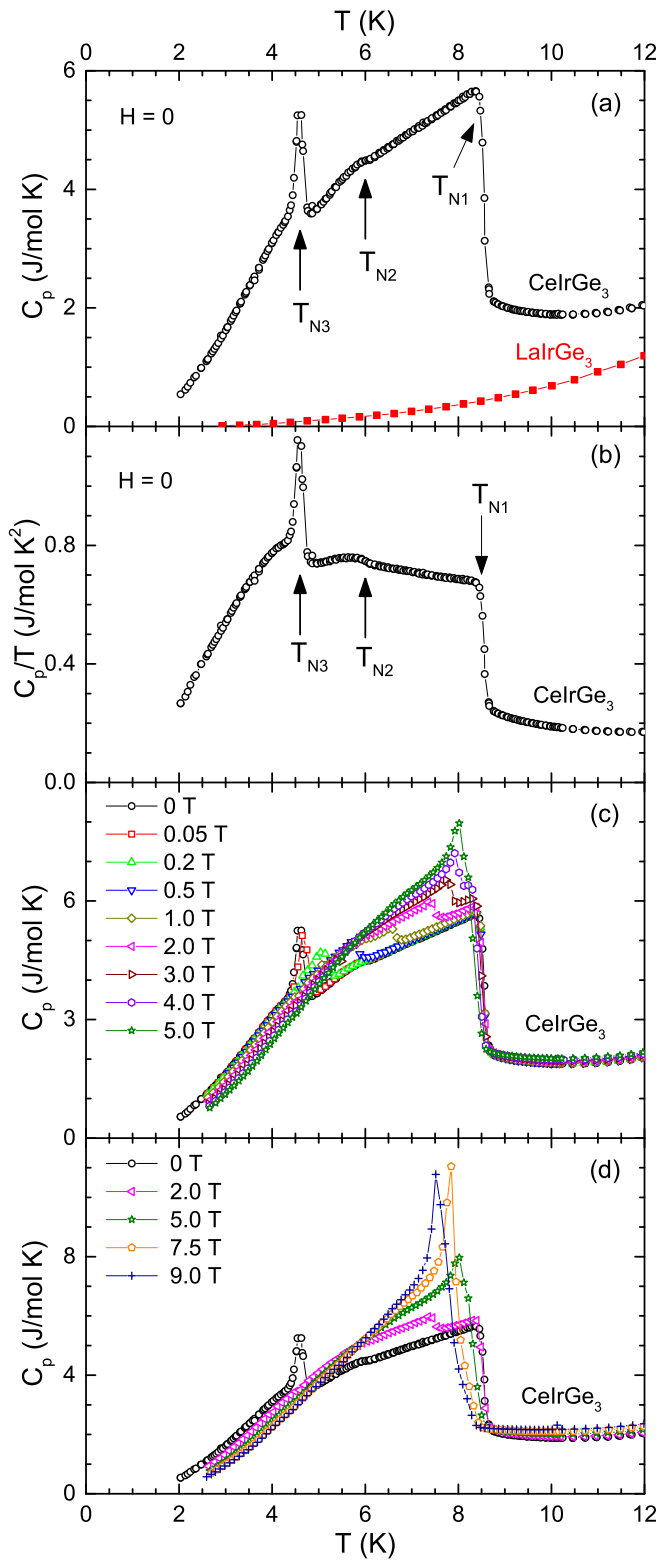}
\caption {(Color online) (a) Temperature $T$ dependence of low-$T$ heat capacity $C_{\rm p}$ of CeIrGe$_{3}$ and LaIrGe$_{3}$ measured in magnetic field $H=0$ for $2 \leq T \leq 12$~K\@. (b) $C_{\rm p}/T$ versus $T$ plot for $H=0$ data of CeIrGe$_{3}$. (c) and (d) $C_{\rm p}(T)$ data of CeIrGe$_{3}$ measured at different $H$ for $H \leq 9$~T. The arrows in (a) and (b) mark the transitions at $T_{\rm N1}= 8.5$~K, $T_{\rm N2}= 6.0$~K  and $T_{\rm N3}= 4.6$~K\@. }
\label{fig:HC}
\end{figure}

\begin{figure}
\includegraphics[width=\columnwidth]{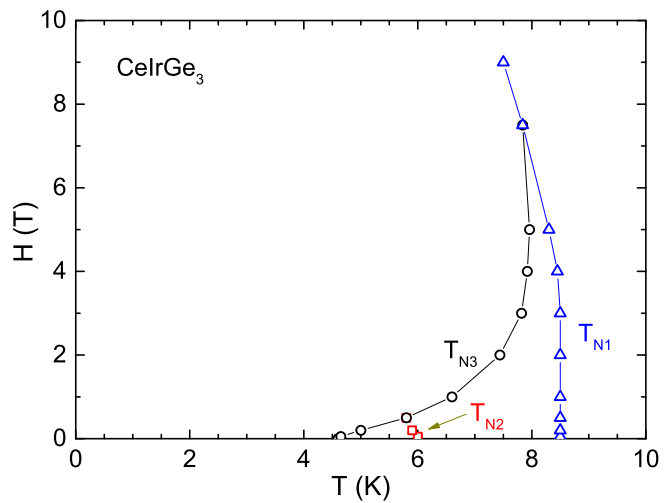}
\caption {(Color online) $H$-$T$ phase diagram for the Ce magnetic ordering in CeIrGe$_{3}$ determined from the heat capacity $C_{\rm p}(T, H)$ data in Fig.~\ref{fig:HC}. }
\label{fig:H-T}
\end{figure}

Figure~\ref{fig:HC} shows the heat capacity $C_{\rm p}(T)$ data of CeIrGe$_{3}$ measured at different $H$ for $H \leq 9$~T\@. The zero-field $C_{\rm p}(T)$ data [Fig.~\ref{fig:HC}(a)] exhibit three anomalies near 4.6~K, 6.0~K and 8.5~K. While the anomalies near 4.6~K and 8.5~K are well pronounced, the 6.0~K anomaly is much weaker. This anomaly can be better visualized in $C_{\rm p}/T$ versus $T$ plot shown in Fig.~\ref{fig:HC}(b). We define the three transitions temperatures $T_{\rm N1}= 8.5$~K, $T_{\rm N2}= 6.0$~K  and $T_{\rm N3}= 4.6$~K\@. Because of the influence of magnetic interaction in the ordered state and crystal field above $T_{\rm N1}$, it is difficult to estimate the Sommerfeld coefficient $\gamma$ precisely. Nevertheless, we estimate  $\gamma = 102(4)$~mJ/mole\,K$^{2}$ by fitting the $C_{\rm p}/T$ versus $T^2$ plot over $12.5~{\rm K} \leq T \leq 14.5$~K (fit not shown) according to $C_{\rm p}/T = \gamma +\beta T^{2}$. The fit also gave $\beta=0.44(3)$~mJ/mole\,K$^{4}$ which in turn gives Debye temperature $\Theta_{\rm D} = 281(6)$~K\@. For LaIrGe$_{3}$ $\gamma$ is found to be 5.9(1)~mJ/mole\,K$^{2}$ and  $\Theta_{\rm D} = 250(2)$~K\@.

We notice that the jump in heat capacity at $T_{\rm N1}$ is much lower than the expected mean field jump of 12.48~J/mol\,K for purely magnetic two-level system (effective $J=1/2$) which could be attributed to the presence of Kondo effect. We estimate the Kondo temperature $T_{\rm K}$ following Besnus {\it et al}. \cite{Besnus} who suggested a universal behavior for jump in magnetic heat capacity $\Delta C_{\rm mag}$ versus $T_{\rm K}/T_{\rm N}$ plot for Ce-based Kondo lattice systems. For CeIrGe$_{3}$, we find $\Delta C_{\rm mag} \approx 5$~J/mol\,K at $T_{\rm N1} = 8.5$~K which corresponds to $T_{\rm K}/T_{\rm N} \approx 1.15$ in that universal plot. This yields $T_{\rm K} \approx  9.8$~K\@. Another estimate of $T_{\rm K}$ follows from the Weiss temperature, $T_{\rm K} \approx |\theta_{\rm p}|/4.5 $ \cite{Gruner}, which for $\theta_{\rm p} = -31$~K gives $T_{\rm K} \approx  6.9$~K\@ which is little lower than above estimate of $T_{\rm K} \approx  9.8$~K\@.

The $C_{\rm p}(T)$ data measured at different $H$ shown in Figs.~\ref{fig:HC}(c) and \ref{fig:HC}(d) present an interesting behavior. The temperature of the anomaly at $T_{\rm N3}$ increases with increasing field until it merges with $T_{\rm N1}$ at around 7.5~T\@. The anomaly at $T_{\rm N2}$ shifts downwards and is not detectable for fields above 0.5~T\@.  On the other hand the anomaly at $T_{\rm N1}$ is almost insensitive to fields for $H\leq 4$~T, above this a weak decrease is observed in $T_{\rm N1}$ on increasing $H$. The $H$--$T$ phase diagram obtained from $C_{\rm p}(T, H)$ data is shown in Fig.~\ref{fig:H-T}. The $H$--$T$ phase diagram clearly shows the $T$ dependence of $T_{\rm N1}$, $T_{\rm N2}$ and $T_{\rm N3}$ and depicts a complex magnetic behavior of CeIrGe$_{3}$. 

Furthermore, we see that the peak height of $C_{\rm p}(T)$ anomaly at $T_{\rm N3}$ initially decreases up to 1~T, then increases at $H\geq 2$~T\@. The peak height of $C_{\rm p}(T)$ anomaly at $T_{\rm N1}$ initially increases very slowly up to 4~T above which a rapid increase is observed, particularly at $H =7.5$~T and 9~T\@. We thus see that the jump in heat capacity increases very rapidly at $H \geq7.5$~T\@. An increase in $\Delta C_{\rm mag}$ would suugest that the  Kondo interaction and hence $c$-$f$ hybridization weakens with increasing field and the system moves towards a more localized state. A change in the nature of phase transition from second order to first order can also lead to an increase in the heat capacity jump in high field limit.

\begin{figure}
\includegraphics[width=\columnwidth]{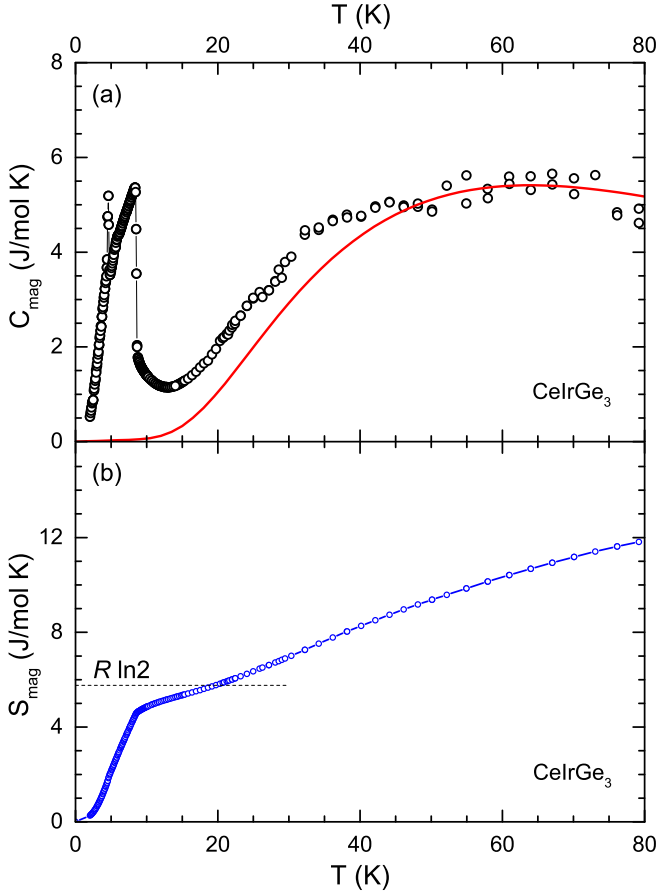}
\caption {(Color online) (a) Temperature $T$ dependence of magnetic heat capacity $C_{\rm mag}$  of CeIrGe$_{3}$. The solid curve represents the crystal electric field contribution to heat capacity according to the crystal field level scheme deduced from the inelastic neutron scattering data including a $\gamma T$ contribution with $\gamma = 5$~mJ/mole\,K$^{2}$.   (b) Magnetic entropy $S_{\rm mag} $ versus $T$. }
\label{fig:HCmag}
\end{figure}

The magnetic contribution to heat capacity $C_{\rm mag}(T)$ is shown in Fig.~\ref{fig:HCmag}(a). The $C_{\rm mag}(T)$ was estimated by subtracting off the lattice contribution using the heat capacity data of isostructural LaIrGe$_{3}$. We see a broad Schottky-type anomaly (centered around 60~K) in $C_{\rm mag}(T)$ which can be attributed to crystal field. A comparison of $C_{\rm mag}(T)$ data with the CEF contribution to heat capacity $C_{\rm CEF}(T)$ estimated according to the CEF level scheme obtained from the analysis of inelastic neutron scattering data in Sec.~\ref{Sec:INS} is presented in Fig.~\ref{fig:HCmag}(a) (solid red curve). A very reasonable agreement is observed between the $C_{\rm mag}(T)$ and $C_{\rm CEF}(T)$ in reproducing the Schottky-type feature. The magnetic contribution to entropy $S_{\rm mag}(T)$ obtained by integrating the $C_{\rm mag}(T)/T$ versus $T$ plot is shown in Fig.~\ref{fig:HCmag}(b). The $S_{\rm mag}(T)$ shows that $S_{\rm mag}$ attains a value of $\approx 80\%$ of $R\ln2$ at $T_{\rm N1}$. The magnetic entropy of $R\ln2$ is achieved by 20~K\@. 

The reduced magnetic entropy at $T_{\rm N1}$ again indicates sizable Kondo effect in this compound. The magnetic entropy of magnetically ordered Kondo lattice system in a simple two-level model with an splitting energy of $k_{\rm B}T_{\rm K}$ is given by \cite{Yashima}
\begin{equation}
S_{\rm mag} = \ln(1+e^{-T_{\rm K}/T_{\rm N}}) + \frac{T_{\rm K}}{T_{\rm N}}\left(\frac{e^{-T_{\rm K}/T_{\rm N}}}{1+e^{-T_{\rm K}/T_{\rm N}}}\right).
\end{equation}
For $S_{\rm mag}$ of $\approx 80\%$ of $R\ln2$ at $T_{\rm N1}$ this suggests $T_{\rm K} \approx  9.5$~K, in good agreement with above estimate of $T_{\rm K} \approx  9.8$~K from $\Delta C_{\rm mag}$.

\begin{figure} 
\includegraphics[width=\columnwidth]{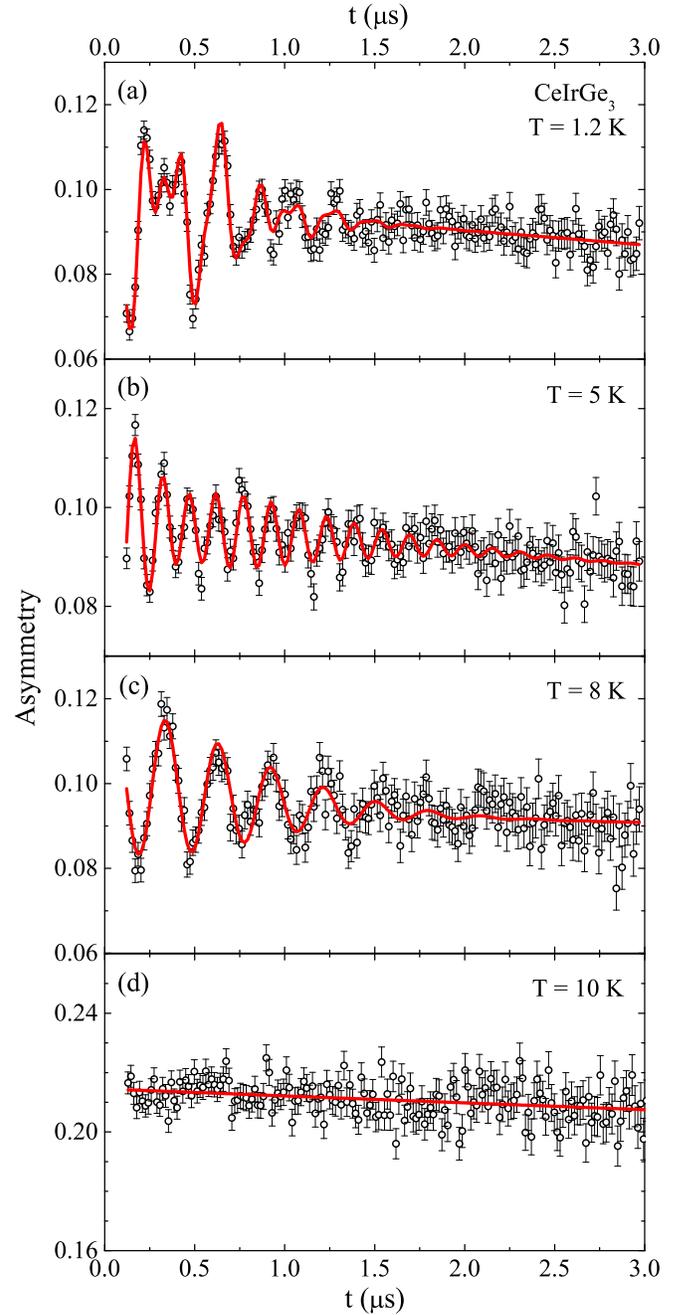}
\caption{\label{fig:MuSR1} (Color online) Zero field muon-spin asymmetry function $G_z$ versus time $t$, $\mu$SR spectra of CeIrGe$_{3}$ for indicated representative temperatures. The solid lines represent the fits to the data by the function described in the text.}
\end{figure}

\section{Muon Spin relaxation}

\begin{figure} 
\includegraphics[width=\columnwidth]{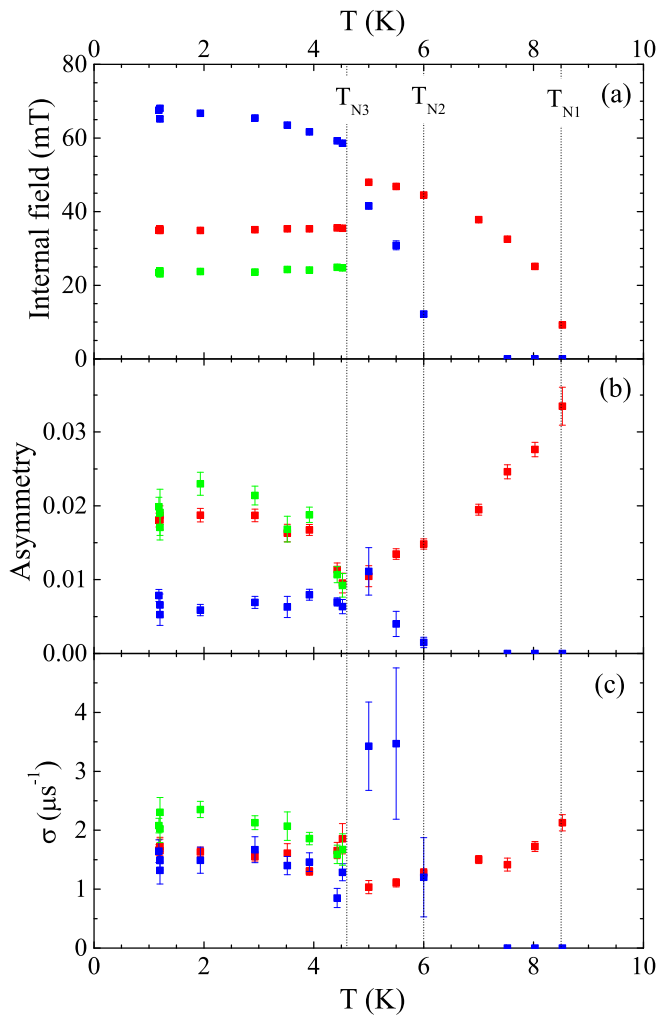}
\caption{\label{fig:MuSR2} (Color online) Temperature $T$ dependence of (a) the internal field $H_{{\rm int},i}$ (b) the initial asymmetries $A_i$, and (c) Gaussian decay rate $\sigma_i$ obtained from the analysis of zero field $\mu$SR spectra of CeIrGe$_{3}$ collected at various temperatures. The vertical dotted lines mark the three transitions at $T_{\rm N1}= 8.5$~K, $T_{\rm N2}= 6.0$~K  and $T_{\rm N3}= 4.6$~K\@.}
\end{figure}

The representative zero-field $\mu$SR spectra are shown in Fig.~\ref{fig:MuSR1} for $T=1.2,~5$, 8 and 10~K\@. At $T> T_{\rm N1}$ (10~K) the $\mu$SR spectra are slow depolarizing, described by a simple exponential decay arising from spin fluctuations. However, clear oscillations are observed in the $\mu$SR spectra at $T \leq T_{\rm N1}$, accompanied with a loss in initial asymmetry (see Fig.~\ref{fig:MuSR1}). This is a classic signature of long range magnetic ordering. The ordered state $\mu$SR spectra are well described by a combination of oscillatory functions convoluted with a Gaussian envelope,  and depending on the temperature ranges data could be fitted with either one, two or three oscillating functions. The fitting function that we used to fit the ordered state $\mu$SR spectra is 
\begin{equation}
 G_z(t)= \sum\limits_{i=1}^3 A_i \cos(\omega_i t + \phi) \,{\rm e}^{-(\sigma_i t)^2/2} + A_0 \,{\rm e}^{-\lambda t}+ A_{\rm BG}.
\end{equation}
Here $A_i$ and $A_0$ are the initial asymmetries of oscillatory and exponential components, $\sigma_i$ are the muon depolarisation rates (arising from a distribution of internal fields) forming a Gaussian envelope to the oscillating component with a frequency of $\omega_i$ and a phase $\phi$, $\lambda$ is the muon depolarization rate and $A_{\rm BG}$ is the background. The value of $A_{\rm BG}$ was estimated by fitting spectra at 10~K and was kept fix for fitting other temperature points. The frequency $\omega = \gamma_\mu H_{\rm int}$, where $\gamma_\mu = 2\pi \times 135.53$~MHz/T is the muon gyromagnetic ratio and $H_{\rm int}$ is the internal field at the muon site.

In order to fit the $\mu$SR data three oscillating functions are required for $T\leq 4.5$~K, two oscillating functions are required for $4.5< T\leq 6$~K and only one is required for $6< T\leq 8.5$~K. The fit of the representative spectra are shown by solid lines Fig.~\ref{fig:MuSR1}. The temperature dependences of the fitting parameters are shown in Fig.~\ref{fig:MuSR2}. It is evident from Fig.~\ref{fig:MuSR2}(a) that muons sense three different internal fields corresponding to three different frequencies. Each change in the number of frequencies corresponds to an anomaly in the heat capacity presented in previous section.  The $T$ dependent initial asymmetries of the oscillatory component show that CeIrGe$_{3}$ is fully ordered below T$_{N1 }$ [see Fig.~\ref{fig:MuSR2}(b)]. The damping of the oscillating component $\sigma$ also shows an interesting temperature dependence [see Fig.~\ref{fig:MuSR2}(c)]. The $\sigma$ is more or less temperature independent over a wide range of temperature with the exception of the intermediate $T$ regime where there is a rapid increase in $\sigma$. This implies that there is a broad distribution of internal fields.

\section{Neutron Diffraction}

\begin{figure}
\includegraphics[width=\columnwidth]{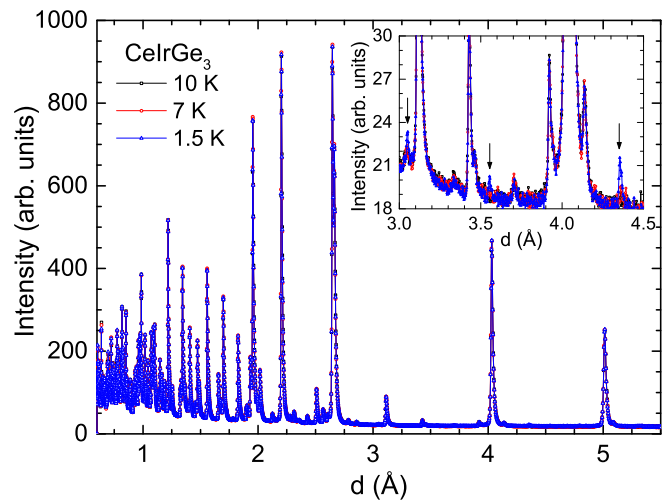}
\caption{\label{fig:ND1} (Color online)  Comparison of neutron diffraction (ND) patterns recorded at 1.5~K, 7~K and 10 K\@.  Inset: An expanded scale plot of ND patterns showing the magnetic Bragg peaks as indicated by arrows.}
\end{figure}

In order to get further insight into the magnetic properties of CeIrGe$_3$ and discern the nature of the low-temperature phase transitions we performed the neutron diffraction measurements. The ND data collected at 1.5~K, 7~K and 10~K are shown in Fig.~\ref{fig:ND1}. The 10~K ND data were successfully refined using the BaNiSn$_3$-type non-centrosymmetric tetragonal ($I4\,mm$) structural model similar to the isostructural compound CeCoGe$_3$ \cite{Smidman2013}. The expanded scale plot shown in the inset of Fig.~\ref{fig:ND1} clearly shows the appearance of additional reflections at 1.5~K and 7~K. The peaks are observed only at low momentum-transfer region, revealing the magnetic origin of these reflections.  The intensity of these magnetic Bragg peaks is temperature dependent. The additional weak reflections at 7~K can be accounted with an incommensurate propagation vector {\bf k} = (0,0,0.688(3)). The absence of zero satellites strongly indicates that the magnetic ordering is a longitudinal spin density wave and a quantitative magnetic structure refinement confirmed this conclusion. 

\begin{figure}
\includegraphics[width=\columnwidth]{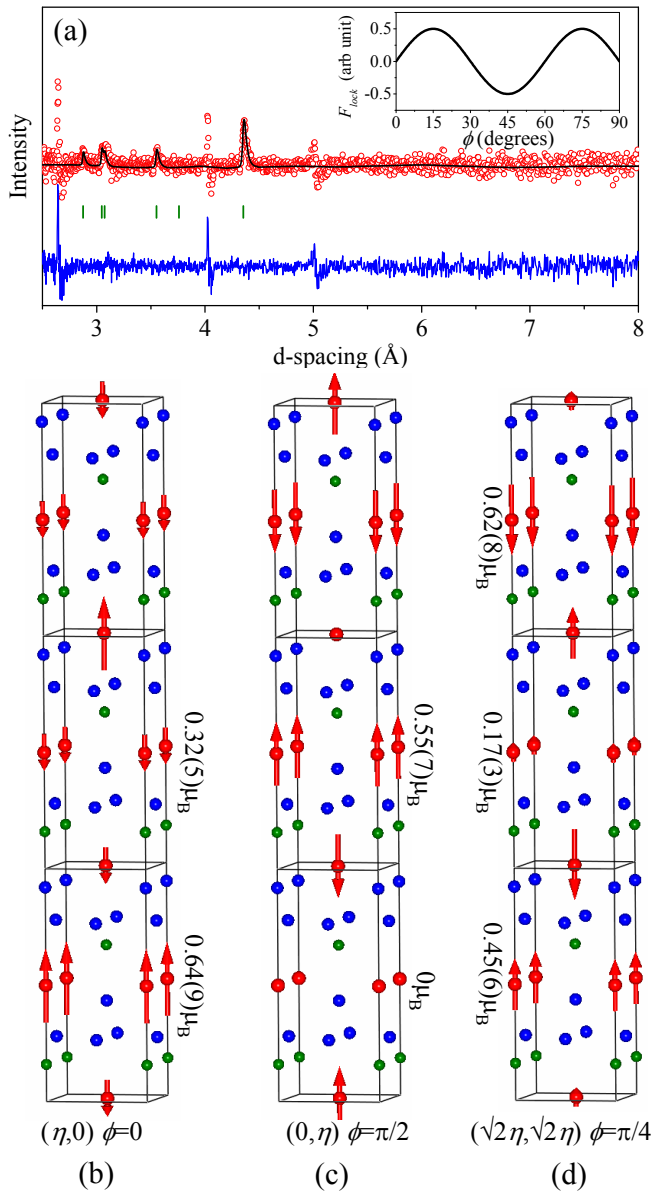}
\caption{\label{fig:Mag_struct} (Color online) (a) Rietveld refinement of the magnetic intensity obtained as a difference between the neutron diffraction patterns collected at $T=1.5~K$ and $T=10$~K ($R_{\rm mag}=8.56\%$). Inset: Magnetic phase dependence of the lock-in free energy invariant, $F_{\rm lock}$. Commensurate magnetic structures at 1.5~K corresponding to (b) $\phi = 0$, (c) $\phi = \pi/2$, and (d) $\phi = \pi/4$. For all cases, the distinct values of the ordered moments are shown (the refined amplitude of the commensurate spin density wave is $0.64(9)\,\mu_{\rm B}$). The arrows denote the ordered Ce$^{3+}$ magnetic moment directions.}
\end{figure}

The diffraction pattern measured at $T=1.5$~K qualitatively looks very similar to the 7~K data set. However, the refinement procedure [Fig.~\ref{fig:Mag_struct}(a)] yields a commensurate propagation vector {\bf k} = (0,0,2/3) within the error bar of the fitting ($k_z=0.667(1)$). This key result indicates that the anomaly found in the specific heat and susceptibility data at $T_{\rm N3}=4.6$~K should be assigned to a magnetic lock-in transition. A further support of this scenario comes from symmetry arguments \cite{Stokes,Campbell2006}. The longitudinal spin density wave associated with the (0,0,$k_z$)-line of symmetry transforms as a two-dimensional time-odd irreducible representation mLD4LE4($\eta_1$,$\eta_2$) of the $I4\,mm$ space group. The symmetry of this magnetic order parameter impels the presence of six-power lock-in invariant in the Landau free-energy decomposition: $F_{\rm lock} = 3\eta_1^5 \eta_2 - 10\eta_1^3\eta_2^3 + 3\eta_1\eta_2^5$. The invariant is allowed only at $k_z=2/3$ and an activation of this energy term naturally explains the transition. Another important point is that at the commensurate value of $k_z=2/3$, the Landau free energy also allows a term which couples a homogeneous ferromagnetic component $m_z$ along the fourfold axis: $m_z(\eta_1^3 - 3\eta_1\eta_2^2$) and $m_z(\eta_2^3 - 3\eta_1^2\eta_2$). This is in excellent agreement with the magnetization data revealing the net moment below $T_{\rm N3}$ (see Figs.~\ref{fig:Chi-LowH} and \ref{fig:MH}). In the temperature range of $T_{\rm N3} < T < T_{\rm N1}$, the phase with the spontaneous magnetization can be induced by magnetic field (metamagnetic behavior), which implies a field induced lock-in transition.

The commensurate value of the propagation vector also implies that the ordered moments localized on the Ce sites depend on the global phase $\phi$ of the magnetic structure. It is well known, however, that the magnetic structure factors are insensitive to $\phi$ and therefore the magnetic structure cannot be unambiguously determined directly from the neutron diffraction data. The problem can be overcome, if we consider the lock-in term as a function of the magnetic phase. Different values of $\phi$ correspond to different directions of the magnetic order parameter in the ($\eta_1$,$\eta_2$) representation space. $F_{\rm lock}$ depends on the magnetic global phase and takes extremal values at some particular values of $\phi$ (inset of [Fig.~\ref{fig:Mag_struct}(a)]). At these values the proximity of the magnetic structure to be locked is maximal. Note, $F_{\rm lock}$ vanishes at $\phi = 0$ and $\phi = \pi/2$ and therefore the corresponding magnetic structures [shown in Fig.~\ref{fig:Mag_struct}(b) and (c)] cannot be locked. The values of $\phi $ which maximize (minimize) the lock-in term correspond to equivalent magnetic structures, and the case of $\phi = \pi/4$ is shown in Fig.~\ref{fig:Mag_struct}(d). Thus, the presented above symmetry-based approach allows to unambiguously determine the magnetic structure of CeIrGe$_3$  and can be efficiently used for other systems. It is seen that the ordered moments are aligned along the $c$-axis and the commensurate spin density wave magnetic structure consists of an alternation of Ce-layers with the ordered moments $0.62(8)\,\mu_{\rm B}$, $0.45(6)\,\mu_{\rm B}$ and $0.17(3)\,\mu_{\rm B}$ stacked along the $c$-direction. It should be noted that the observation of the three frequencies in our $\mu$SR study [see Fig.~\ref{fig:MuSR2}(a)] is consistent with the proposed commensurate magnetic structure having three types of the layers with distinct Ce moments.

\section{\label{Sec:INS} Inelastic neutron study}

\begin{figure} 
\includegraphics[width=\columnwidth]{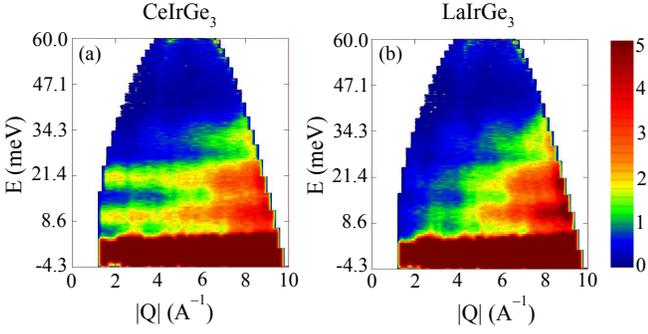}
\caption{\label{fig:INS-contour} (Color online) Inelastic neutron scattering response, a color-coded contour map of the intensity, energy transfer $E$ versus momentum transfer $|Q|$ for (a) CeIrGe$_{3}$ and (b) LaIrGe$_{3}$ measured at 2~K with the incident energy $E_{i} = 67.6$~meV.}
\end{figure}

\begin{figure}
\includegraphics[width=8cm]{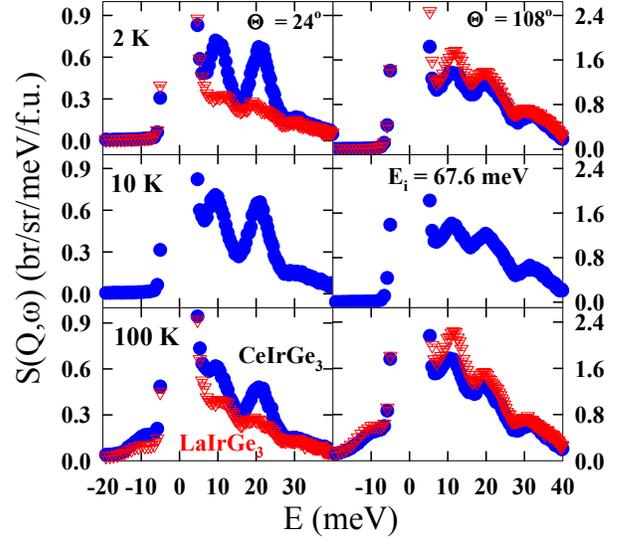}
\caption{\label{fig:INS-S} (Color online) $\Theta$-integrated inelastic scattering intensity $S(Q,\omega)$ versus energy transfer $E$ for CeIrGe$_{3}$ and LaIrGe$_{3}$ at $\Theta = 24^\circ$ (left panels) and $\Theta = 108^\circ$ (right panels) measured with $E_{i}= 67.6$~meV at 2~K, 10~K and 100~K on IN4.}
\end{figure}

The INS scattering responses from CeIrGe$_{3}$ and LaIrGe$_{3}$ measured with $E_{i} =67.6$~meV at $T= 2$~K are shown in Fig.~\ref{fig:INS-contour} as the color coded intensity maps. A comparison of the INS scattering responses from CeIrGe$_{3}$ and LaIrGe$_{3}$ clearly reveals two excitations (near 9.7 and 20.9~meV) of magnetic origin for CeIrGe$_{3}$. 
The scattering angle $\Theta $-integrated one-dimensional (1D) energy cuts of INS responses at 1.2, 10 and 100~K for low ($\langle \Theta \rangle = 24^\circ$) and high ($\langle \Theta \rangle = 108^\circ$) angles are shown in Fig.~\ref{fig:INS-S}. The two magnetic excitations are very clear at low $\Theta$ (left panels, Fig.~\ref{fig:INS-S}).   At high $\Theta$ (right panels, Fig.~\ref{fig:INS-S}) both La and Ce show similar excitations suggesting that these excitations are mainly of phononic origin. 

\begin{figure}
\includegraphics[width=\columnwidth]{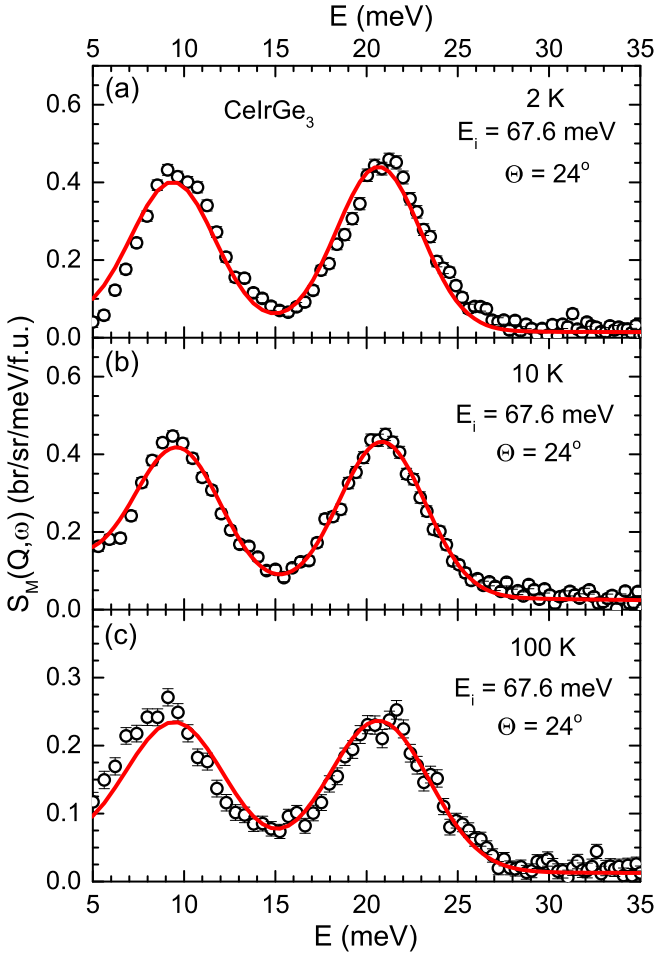}
\caption{\label{fig:INS-Smag} (Color online) $\Theta$-integrated inelastic magnetic scattering intensity $S_{\rm M}(Q,\omega)$ versus energy transfer $E$ for CeIrGe$_{3}$ at $\Theta = 24^\circ$ measured with $E_{i}$ = 67.6~meV at (a) 2~K, (b) 10~K and (c) 100~K\@. The solid lines are the fits of the data based on CEF model. }
\end{figure}

The magnetic scattering $S_{\rm M}(Q,\omega)$ for CeIrGe$_{3}$ is shown in Fig.~\ref{fig:INS-Smag} which was obtained after subtracting the phonon background using the INS data of LaIrGe$_{3}$. The $S_{\rm M}(Q,\omega) = S(Q,\omega)_{\rm CeIrGe_3} - \alpha\,S(Q,\omega)_{\rm LaIrGe_3}$ where $ \alpha = 0.86$ is the ratio of neutron scattering cross sections of CeIrGe$_3$ and LaIrGe$_{3}$. The two magnetic excitations near 9.7 and 20.9~meV seen in Fig.~\ref{fig:INS-Smag} are well accounted by a model based on crystal electric field. The six-fold degenerate ground state of Ce$^{3+}$ splits into three doublets in tetragonal symmetry environment. The crystal field Hamiltonian for the tetragonal symmetry (point symmetry
$C_{4v}$) Ce$^{3+}$ atoms are given by
\begin{equation}\label{H-CEF}
 H_{\rm Tetra} = B_{2}^{0}O_{2}^{0} + B_{4}^{0}O_{4}^{0} + B_{4}^{4}O_{4}^{4}
\end{equation}
where $B_{n}^{m}$ are CEF parameters and $O_{n}^{m}$ are the Stevens operators.

\begin{figure}[h]
\includegraphics[width=\columnwidth, keepaspectratio]{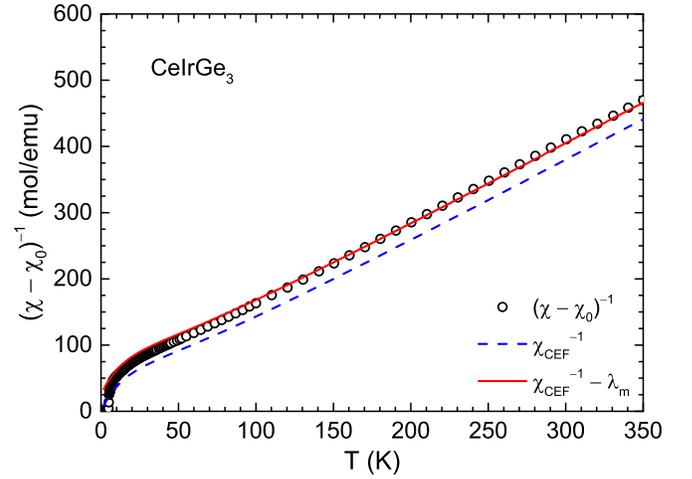}
\caption {(Color online) Inverse zero-field-cooled dc magnetic susceptibility of CeIrGe$_{3}$ after subtracting the temperature $T$-independent contribution $(\chi - \chi_0)^{-1}$ as a function of $T$. The dashed blue line is the inverse of crystal electic field susceptibility $\chi_{\rm CEF}^{-1}$ corresponding to the CEF parameters obtained from the analysis of inelastic neutron scattering data. The solid red line represnts the CEF susceptibility after including the molecular field constant $\lambda_{\rm m}$.}
\label{fig:Chi-CEF}
\end{figure}

In order to obtain a unique set of CEF parameters we fitted the 2~K, 10~K and 100~K $S_{\rm M}(Q,\omega)$ data simultaneously. The CEF parameters obtained from the analysis of INS data are listed in Table~\ref{tab:CEF} and the fits of INS data are shown by solid red curves in Fig.~\ref{fig:INS-Smag}. A very small difference between the fitted line and the data at 1.2~K is due to the fact that we have not included the molecular field term in our calculation. Further small difference at 100~K is attributed to the thermal expansion, which will also change the CEF potential. The first excited doublet is found to be situated at 112.7~K (9.71~meV) and the second excited doublet at  242.4~K (20.89~meV) with respect to ground state doublet. The CEF wave functions obtained are:
\begin{equation}\label{Eq:wavefn}
\begin{split}
\Psi_0 & =   (0.8826) |\pm \frac{3}{2}\rangle -  (0.4702) |\mp \frac{5}{2}\rangle \\
\Psi_1  & =    |\pm \frac{1}{2}\rangle \\
\Psi_2  & =   (0.4702) |\pm \frac{3}{2}\rangle +  (0.8826) |\mp \frac{5}{2}\rangle,
\end{split}
\end{equation}
where $\Psi_0 $ corresponds to ground state doublet, $\Psi_1 $ to first excited state doublet and $\Psi_2 $ to second excited state doublet.

\begin{table} [b]
\caption{\label{tab:CEF} Crystal field parameters $B_{n}^{m}$ and splitting energies $\Delta_i$ of excited states (with respect to ground state, $\Delta_0 \equiv 0$) obtained from the analysis of the inelastic neutron scattering data of CeIrGe$_{3}$.}
\begin{ruledtabular}
\begin{tabular}{ccccc}
$B_2^0$  & $B_4^0$  & $B_4^4$  & $\Delta_1$  & $\Delta_2$ \\
(meV) & (meV)  &  (meV) &  (meV) & (meV) \\
\hline
$0.451(2)$ & $0.026(1)$ & 0.323(6) & 9.71  & 20.89 \\
\end{tabular}
\end{ruledtabular}
\end{table}

The CEF contribution to specific heat $C_{\rm CEF}(T)$ estimated according to the obtained CEF level scheme is shown by the solid red curve in Fig.~\ref{fig:HCmag} which shows a very reasonable agreement with the experimental $C_{\rm mag}(T)$ data. A comparison of the CEF susceptibility $\chi_{\rm CEF}(T)$ with the dc susceptibility after subtracting the $T$-independent contribution $(\chi - \chi_0)(T)$ is shown in Fig.~\ref{fig:Chi-CEF}. Taking into account the molecular field constant $\lambda_{\rm m}$, the susceptibility is given by $(\chi - \chi_0)^{-1} = \chi_{\rm CEF}^{-1} - \lambda_{\rm m}$. As shown by the solid red line, a very good agreement is observed between the $(\chi - \chi_0)^{-1}$ data and ($\chi_{\rm CEF}^{-1} - \lambda_{\rm m}$) for $\lambda_{\rm m} = -25(2)$~mol/emu.

We estimate the ground state magnetic moment using the relation,
\begin{equation}\label{Eq:moment}
\begin{split}
\langle \mu_x \rangle & =  \langle \Psi_0| \frac{g_J}{2}(J^+ + J^-) |\Psi_0 \rangle \\
\langle \mu_z \rangle & =  \langle \Psi_0| {g_J}(J_z) |\Psi_0 \rangle 
\end{split}
\end{equation}
which gives the $ab$-plane moment $\langle \mu_x \rangle = 0.80~\mu_{\rm B}$ and $c$-direction moment $\langle \mu_z \rangle = 0.53~\mu_{\rm B}$. The positive $B_2^0$ suggests the moment to be in $ab$-plane which is different from the observed moment direction (along $c$-axis) from the ND data. This indicates that the anisotropic exchange interactions are playing an important role, dominating over the single-ion CEF anisotropy, in determining the moment direction. A very similar situation has been observed for CeRhGe$_{3}$ \cite{Hillier2012}, CeCuAl$_{3}$ \cite{Adroja2012b}, CeRu$_{2}$Al$_{10}$ \cite{Khalyavin2010, Kato2011,Strigari2012, Bhattacharyya2014} and CeOs$_{2}$Al$_{10}$ \cite{Kato2011,Strigari2013,Adroja2016}, where the direction of the ordered moments is different from that expected from the single-ion CEF anisotropy. In the case of CeRhGe$_{3}$  the single-ion CEF anisotropy predicts the moments to lie in the $ab$-plane, however, the ND revealed the ordered moments to be directed along the $c$-axis \cite{Hillier2012}. For CeCuAl$_{3}$ the CEF predicts $a$-axis to be the direction of moments, whereas the ND finds ordered moments oriented along the $c$-axis \cite{Adroja2012b}. On the other hand, for CeCoGe$_{3}$ the CEF prediction of moments along the $c$-axis is found to be consistent with that determined from ND study \cite{Smidman2013}. For CeAuAl$_{3}$ also the CEF predicted direction of moments ($ab$-plane) is found to agree with that determined by ND \cite{Adroja2015}. 

\begin{figure} 
\includegraphics[width=\columnwidth]{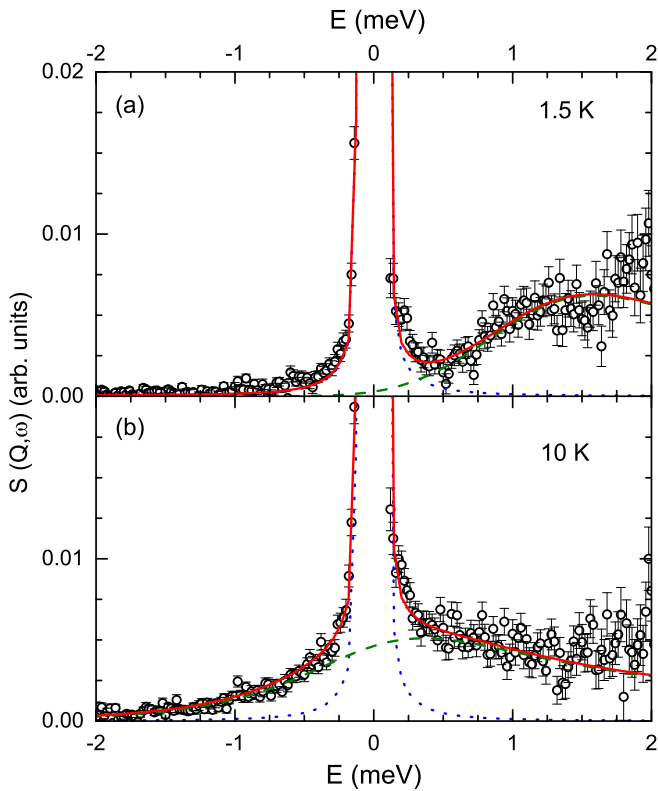}
\caption{\label{fig:INS-LE} (Color online) Low energy inelastic neutron scattering for CeIrGe$_{3}$  measured at (a) 1.5~K and (b) 10~K with the incident energy $E_{i} = 3.1$~meV. The solid line is the fit with a Lorentzian line-shape function for the quasielastic and inelastic components. The dashed (elastic peak) and dashed (quasielastic and inelastic peaks) lines represent the components of the fit. }
\end{figure}

In order to have an idea about the spin-wave energy scale and estimate the Kondo temperature we also performed the low-energy INS measurements with $E_{i} = 3.1$~meV which are shown in Fig.~\ref{fig:INS-LE}.
The 1D cuts in Fig.~\ref{fig:INS-LE} display the total scattering summed over all scattering angles from $10^\circ$  to $ 115 ^\circ$. At 1.5 K, the scattering of spin-wave origin is clearly seen at energies above 0.8~meV. Further, we also see a reminiscent  of a gapped spin wave behavior with a possible energy gap of $\sim 2$~meV. In the paramagnetic state, at 10~K we clearly see a significant contribution from quasielastic scattering. We have fitted the low energy INS data using a Lorentzian line-shape function for the quasielastic and inelastic components, the fits are shown in Fig.~\ref{fig:INS-LE} along with the components. From the quasielastic linewidth at 10~K we estimate $T_{\rm K} \approx  12.8(8)$~K which is in very good agreement with that estimated from the heat capacity data in Sec.~\ref{Sec:HC}.

\section{Conclusions}

A comprehensive study of magnetic properties of pressure induced noncentrosymmetric heavy-fermion superconductor CeIrGe$_3$ have been performed using $\chi_{\rm ac}(T)$, $\chi(T)$, $M(H)$, $C_{\rm p}(T,H)$,  $\mu$SR, powder ND and INS measurements. In addition to the previously reported magnetic transitions at 8.7~K and 4.7~K, we found evidence for an additional magnetic phase transition near 6~K in our $\chi_{\rm ac}(T)$, $\chi(T)$ and $C_{\rm p}(T)$ measurements. Further confirmation of three magnetic transitions above 2~K comes from our $\mu$SR study.  The oscillatory $\mu$SR asymmetry evidences three transitions at $T_{\rm N1}= 8.5$~K, $T_{\rm N2}= 6$~K  and $T_{\rm N3}= 4.6$~K revealed by different number of oscillating functions (up to three frequencies) for describing the $\mu$SR spectra.  We found the oscillatory $\mu$SR asymmetry to have one frequency for $T_{\rm N2} < T\leq T_{\rm N1}$, two frequencies for $T_{\rm N3} < T\leq T_{\rm N2}$ and three frequencies for $T\leq T_{\rm N3}$ revealing that muons sense different internal fields in these temperature ranges of ordered state. Similar complex magnetic ground states were inferred from the $\mu$SR study on the isostructural compounds  CeRhGe$_{3}$ \cite{Hillier2012} and CeCoGe$_{3}$ \cite{Smidman2013}.

The ND data showed the appearance of weak magnetic Bragg peaks at 7~K and 1.5~K confirming the antiferromagnetic phase  transitions. At 7~K the refinement of ND data reveal an incommensurate magnetic structure, well represented by propagation vector {\bf k} = (0,0,0.688(3)). On the other hand, the magnetic Bragg peaks at 1.5~K are well indexed by commensurate propagation vector {\bf k} = (0,\,0,\,2/3). The magnetic structures in both the high-temperature incommensurate and low-temperature commensurate phases are longitudinal spin density waves with strongly reduced values of the ordered moments. The latter phase couples by symmetry a macroscopic ferromagnetic component, resulting in a strong dependence of the lock-in transition temperature on external magnetic field (metamagnetic behaviour). The global magnetic phase, imposed by the lock-in free-energy invariant to be $\pi/4$ in the commensurate spin density wave, implies an alternation of Ce-layers with the ordered moments $0.62(8)\,\mu_{\rm B}$, $0.45(6)\,\mu_{\rm B}$ and $0.17(3)\,\mu_{\rm B}$ at $T=1.5~K$, which is in full agreement with the three internal fields or frequencies observed in our $\mu$SR study at low temperatures. 

An estimate of $T_{\rm K} \approx  12.8(8)$~K was obtained from the quasielastic linewdith. The high energy INS revealed two well defined magnetic excitations which were accounted by a model based on crystal field. We have extracted information about the CEF states of Ce$^{3+}$. The CEF-spilt excited doublet states are found to be at 9.7~meV and 20.9~meV above the Kramers doublet ground state. The single-ion CEF anisotropy predicts the moment direction in the $ab$-plane, but the moment direction observed from the ND is along the $c$-axis, indicating that anisotropic magnetic exchange interactions are important for the moment direction. Further investigations of the spin wave in CeIrGe$_{3}$ that will give direct information on the anisotropic exchange interactions would be very interesting.

\begin{acknowledgments}
We thank Dr. M. Smidman, Dr. A. Bhattacharyya and Prof. Geetha Balakrishnan for helpful discussion. DTA and VKA acknowledge financial assistance from CMPC-STFC grant number CMPC-09108.
\end{acknowledgments}

\end{document}